\documentclass[11pt]{article}
\def\rp{\stackrel{\rightarrow}{\partial}}
\def\bowstar{\bowtie\kern-0.8em |~}
\begin{document}
\begin{flushright} 
ANL-HEP-PR-99-132 
\end{flushright}

{\Large
\centerline{GEOMETRICAL EVALUATION OF STAR PRODUCTS}
\centerline{ Cosmas Zachos }}

High Energy Physics Division,
Argonne National Laboratory, Argonne, IL 60439-4815, USA \\
\phantom{.} \qquad\qquad{\sl zachos@hep.anl.gov}      
\begin{abstract}
The geometric picture of the star-product based on its Fourier representation
kernel is utilized in the evaluation of chains of star-products and the 
intuitive appreciation of their associativity and symmetries. Such 
constructions appear even simpler for a variant asymmetric product, and 
carry through for the standard star-product supersymmetrization.
\end {abstract}

\noindent\rule{7in}{0.02in}

{\bf \S 1.} ~Groenewold's noncommutative $\star$-product 
\cite{groen} of phase-space functions $f(x,p)$ and $g(x,p)$
 is the unique associative pseudodifferential deformation \cite{bayen}
of ordinary products:
\begin{equation}
\star \equiv ~e^{i\hbar(\stackrel{\leftarrow }{\partial }_{x}
\stackrel{\rightarrow }{\partial }_{p}-\stackrel{\leftarrow }{\partial }_{p}
\stackrel{\rightarrow }{\partial }_{x})/2}~.
\end{equation}
It is the linchpin of deformation (phase-space) quantization 
\cite{moyal,bayen}, as well as 
applications of matrix models and non-commutative geometry ideas in 
M-physics \cite{natied}. In practice, since it involves exponentials of
derivative operators, it may be evaluated through translation 
of function arguments, 
\begin{equation}
f(x,p) \star g(x,p) = f(x+{i\hbar\over 2}\rp_p ,~ p-{i\hbar\over 2}\rp_x)~ 
g(x,p).
\end{equation}
However, explicit evaluations of long strings of star-products 
in this language frequently appear intractable, unless the phase-space 
functions involved consist of exponentials or simple polynomials \cite{cfz}. 

Baker \cite{baker} has utilized the more practical 
Fourier representation of this product as an integral kernel: 
\begin{equation} 
f\star g={1\over \hbar ^2 \pi^2}\int dp^{\prime}  dp^{\prime\prime}  dx' dx''  
~f(x',p')~g(x'',p'')~ \exp \left({-2i\over \hbar} 
\left( p(x'-x'') + p'(x''-x)+p''(x-x') \right )\right) .
\end{equation}
The cyclic determinantal expression multiplying $-2i/\hbar$ in the exponent 
is twice the area of the phase-space triangle 
$({\bf r}'',{\bf r}',{\bf r})$, where ${\bf r}\equiv (x,p)$, namely, 
\begin{equation} 
2A(r'',r',r ) = ({\bf r}' -{\bf r})\wedge ({\bf r}-{\bf r}'')=
{\bf r}''\wedge {\bf r}'   +
{\bf r}'\wedge {\bf r}+
{\bf r}\wedge {\bf r}'' .
\end{equation}
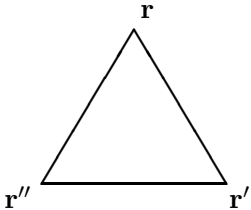
\begin{picture}(75,75)(-30,0)  \thicklines
\put(30,10){\line(1,0){70}}
\put(30,10){\line(3,5){35}}
\put(100,10){\line(-3,5){35}}
\put(70,75){\makebox(0,0)[cc]{{\bf r}}} 
\put(105,5){\makebox(0,0)[cc]{{\bf r}$'$}} 
\put(23,5){\makebox(0,0)[cc]{{\bf r}$''$ }} 
\end{picture}

In this representation, multiple star-products turn out to be simpler to 
evaluate, and the geometrical constructions they motivate exhibit conspicuously 
the symmetries and the associativity of these products. The representation 
thus rises to the level of a `picture', in Dirac's sense of a ``way 
of looking at the fundamental laws which makes their self-consistency obvious"
\cite{pamd}. Such evaluations are illustrated below, with some practical hints,
for the standard star-product, as well as for some common variants and
extensions.

{\bf \S 2.} ~In the Fourier representation, a triple star-product can 
be expressed relatively simply,
\begin{equation} 
f\star g \star h={1\over \hbar ^4 \pi^4}\int \! d\overline{p} dp' dp'' dp''' 
d\overline{x} dx' dx'' dx'''  
f(x',p') g(x'',p'') h(x''',p''') ~\exp {-4i\over \hbar} \left( 
A(r'', r',\overline{r})+A(r''', \overline{r}, r ) \right) ,
\end{equation}
while the intermediary $d\overline{x} ~d\overline{p}$ integrations collapse to 
$\delta$-functions:
\begin{eqnarray}  
f\star g \star h&=&
{1\over \hbar ^2 \pi^2}\! \int \! dp' dp'' dp''' dx' dx'' dx'''  
f(x',p') g(x'',p'') h(x''',p''') ~\times \\
&\times & \delta (x-x'+x''-x''') \delta (p-p'+p''-p''')
~\exp \left({-4i\over \hbar} A(r''', r'',r') \right)    .\nonumber  
\end{eqnarray}  
The product thus amounts to a triangle and a point for the effective 
phase-space argument ${\bf r}=(x,p)$, which lies on the new vertex of 
the parallelogram resulting from doubling up the triangle 
$({\bf r''', r'', r'})$, such that ${\bf r'-r'''}$ is one diagonal. 
The effective argument ${\bf r}$ lies at the end of the {\em other} diagonal, 
starting from ${\bf r''}$,

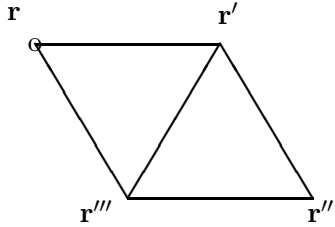
\begin{picture}(200,82)  \thicklines
\put(75,68){\line(1,0){70}}
\put(110,10){\line(1,0){70}}
\put(110,10){\line(3,5){35}}
\put(180,10){\line(-3,5){35}}
\put(110,10){\line(-3,5){35}}
\put(150,80){\makebox(0,0)[cc]{{\bf r}$'$ }} 
\put(185,5){\makebox(0,0)[cc]{{\bf r}$''$ }} 
\put(100,5){\makebox(0,0)[cc]{{\bf r}$'''$ }} 
\put(67,80){\makebox(0,0)[cc]{{\bf r}}} 
\put(75,68){\makebox(0,0)[cc]{o}} 
\end{picture}

It is then straightforward to note how this expression bears no memory of 
the grouping (order of association) in which the two star-multiplications 
were performed, since the vertex  ${\bf r}$ of the parallelogram is reached 
from ${\bf r'''}$
by translating through ${\bf r'-r''}$, or, equivalently, from ${\bf r'}$
by translating through ${\bf r'''-r''}$. As a result, this may well realize 
the briefest  
graphic proof of the distinctive associativity property of the star-product, 
\begin{equation} 
(f\star g) \star h= f\star (g \star h) ~.
\end{equation}

The symmetries of the triple star-product, (1-3 complex conjugacy, 
effective cyclicity, e.t.c.) are now evident by inspection. Moreover, 
integration of this triple product with respect to the effective argument 
$(x,p)$ (tracing), e.g.\ to yield a lagrangian interaction term,  trivially 
eliminates the $\delta$-function to result in a compact cyclic expression 
of the above triangle construction for the three functions star-multiplied,
\begin{equation} 
\int \! dxdp~ f\star g \star h=
{1\over \hbar ^2 \pi^2}\! \int \! dp_1 dp_2 dp_3 dx_1 dx_2 dx_3  
f(x_1,p_1) g(x_2 ,p_2) h(x_3 ,p_3) 
~\exp \left({-4i\over \hbar} A(r_3 , r_2,r_1 ) \right)    .  
\end{equation}

A four function star-product (with three stars) involves the sum 
of the areas of two triangles, $({\bf r}_3,{\bf r}_2,{\bf r}_1$) and 
$({\bf r},{\bf r}_4,{\bf r}_1-{\bf r}_2+{\bf r}_3 )$. 
A five function star-product involves the exponential of the sum of areas of 
two triangles, $({\bf r}_3,{\bf r}_2,{\bf r}_1$) and 
$({\bf r}_5 , {\bf r}_4,{\bf r}_1-{\bf r}_2+{\bf r}_3 )$, with the effective
argument restricted by 
$\delta({\bf r} -{\bf r}_1+{\bf r}_2-{\bf r}_3+{\bf r}_4-{\bf r}_5)\equiv 
\delta({\bf r} -{\bf s}_5 )$. 
Recursively, so on for even numbers of star-multiplied functions,
the phase involving the sums  $A({\bf r}_3,{\bf r}_2,{\bf r}_1)+ 
A({\bf r}_5,{\bf r}_4,{\bf s}_3)+...+A({\bf r},{\bf r}_{2n},{\bf s}_{2n-1})$.
Respectively, for odd numbers of functions,  
the phase involving sums $A({\bf r}_3,{\bf r}_2,{\bf r}_1)+
A({\bf r}_5,{\bf r}_4,{\bf s}_3)+ ... +
A({\bf r}_{2n+1},{\bf r}_{2n},{\bf s}_{2n-1})$, with effective phase-space 
argument restrictions to 
${\bf r}={\bf s}_{2n+1} \equiv \sum_{m=1}^{2n+1} (-)^{m+1} {\bf r}_m$.

As an illustration, consider phase-space points {\bf r}$_i$ arrayed in a 
regular zigzag pattern, 
(i.e.\ for the star-multiplied functions getting support only on those
points on the zigzag). The arguments of the $\delta$-functions, 
{\bf s}$_{2n+1}$, then lie on a line, while the areas of the triangles
demarcated by these points increase in regular arithmetic progression
($A, 2A, 3 A, 4 A,...$):

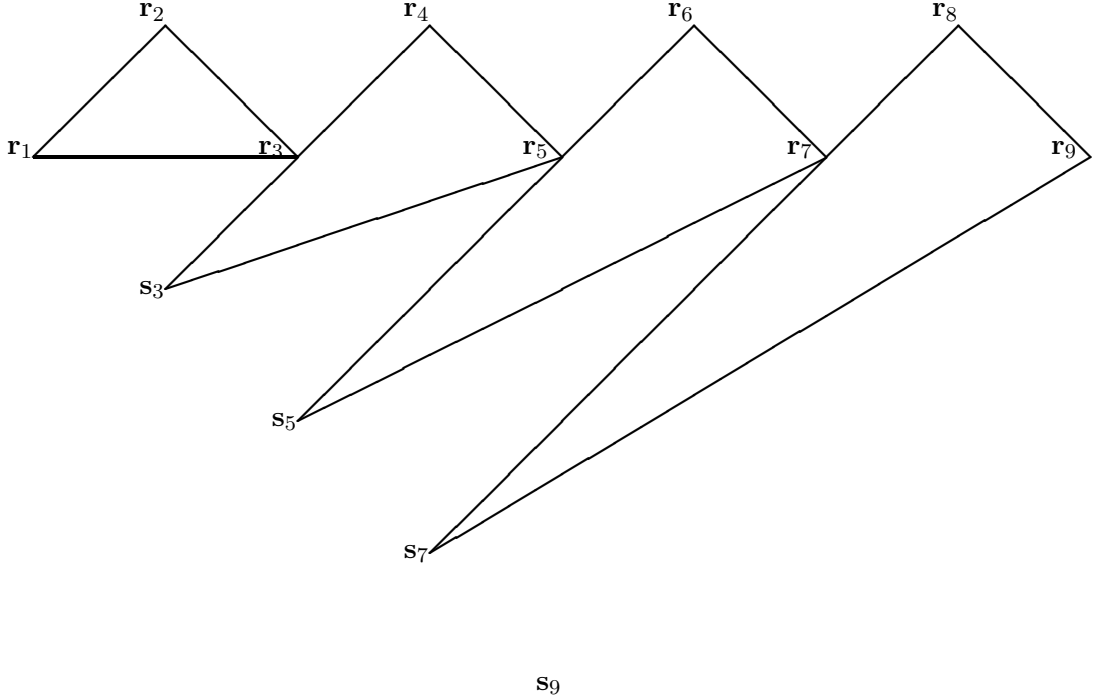
\begin{picture}(410,270)(0,10)  \thicklines
\put(10,210){\line(1,0){100}}
\put(10,210){\line(1,1){50}}
\put(60,260){\line(1,-1){50}}
\put(60,160){\line(3,1){150}}
\put(60,160){\line(1,1){100}}
\put(160,260){\line(1,-1){50}}
\put(110,110){\line(1,1){150}}
\put(110,110){\line(2,1){200}}
\put(260,260){\line(1,-1){50}}
\put(160,60){\line(1,1){200}}
\put(160,60){\line(5,3){250}}
\put(360,260){\line(1,-1){50}}
\put(5,213){\makebox(0,0)[cc]{{\bf r}$_1$}} 
\put(55,265){\makebox(0,0)[cc]{{\bf r}$_2$}} 
\put(100,213){\makebox(0,0)[cc]{{\bf r}$_3$}} 
\put(155,265){\makebox(0,0)[cc]{{\bf r}$_4$}} 
\put(200,213){\makebox(0,0)[cc]{{\bf r}$_5$}} 
\put(255,265){\makebox(0,0)[cc]{{\bf r}$_6$}} 
\put(300,213){\makebox(0,0)[cc]{{\bf r}$_7$}} 
\put(355,265){\makebox(0,0)[cc]{{\bf r}$_8$}} 
\put(400,213){\makebox(0,0)[cc]{{\bf r}$_9$}} 
\put(55,160){\makebox(0,0)[cc]{{\bf s}$_3$}}
\put(105,110){\makebox(0,0)[cc]{{\bf s}$_5$}} 
\put(155,60){\makebox(0,0)[cc]{{\bf s}$_7$}} 
\put(205,10){\makebox(0,0)[cc]{{\bf s}$_9$}} 
\end{picture}

This result is independent of the pitch of the zigzag, i.e.~the angle
at {\bf r}$_2$ ---which, in this figure, is chosen to be $\pi /2$, since this is
a local maximum of the areas $A$ of the triangles for variable pitch but
fixed lengths {\bf r}$_i -${\bf r}$_{i+1}$. One might well wonder if the 
configuration pictured could be used to define a a ``classical path": its 
contribution to the phase of the exponential through the sum of all triangle 
areas, $(1+2+3+4+...) A$, is stationary with respect to 
variations such as this angle variation discussed. The question suggests 
itself, then, whether configurations stationary under {\em all} 
variations can be constructed, leading to a stationary phase evaluation 
of large/infinite star-products, e.g.\ useful in evaluating star-exponentials 
(which yield time-evolution operators in phase-space \cite{bayen}); but, so 
far, no cogent general answers appear at hand. 

{\bf \S 3.} ~A variant of the star-product (cohomologically equivalent to it) 
is the lopsided associative product of Voros \cite{voros}, 
\begin{equation}
\bowstar ~\equiv ~e^{i\hbar \stackrel{\leftarrow }{\partial }_{x}
\stackrel{\rightarrow }{\partial }_{p} }~.  \label{vorosproduct}
\end{equation}

It is sometimes convenient to rotate phase-space variables canonically 
(i.e.\ preserving their Poisson Brackets),
\begin{equation}
(x,p)\mapsto(\frac{x+ip}{\sqrt{-2i}} ~, \frac{x-ip}{\sqrt{-2i}} ) ~,
\end{equation}
to represent this product as
\begin{eqnarray}
\bowstar ~\equiv ~e^{\hbar (\stackrel{\leftarrow }{\partial }_{x}
-i \stackrel{\leftarrow }{\partial }_{p})(\stackrel{\rightarrow }{\partial }_{x}
+i \stackrel{\rightarrow }{\partial }_{p})/2}
&=&e^{i\hbar (\stackrel{\leftarrow }{\partial }_{x}
\stackrel{\rightarrow }{\partial }_{p}-\stackrel{\leftarrow }{\partial }_{p}
\stackrel{\rightarrow }{\partial }_{x})/2}~
e^{i\hbar (\stackrel{\leftarrow }{\partial }_{x}
\stackrel{\rightarrow }{\partial }_{x}+\stackrel{\leftarrow }{\partial }_{p}
\stackrel{\rightarrow }{\partial }_{p})/2} \\
&=&\star ~~e^{-\hbar (\stackrel{\leftarrow }{\partial }_{x}^2+
\stackrel{\leftarrow }{\partial }_{p}^2)/4 }~
e^{-\hbar (\stackrel{\rightarrow }{\partial }_{x}^2 +
\stackrel{\rightarrow }{\partial }_{p}^2)/4}~
e^{\hbar ((\stackrel{\leftarrow }{\partial }_{x}+
\stackrel{\rightarrow }{\partial }_{x})^2 +
(\stackrel{\leftarrow }{\partial }_{p}+
\stackrel{\rightarrow }{\partial }_{p})^2)/4}~.\nonumber
\end{eqnarray}
This turns out to be the covariant transform of the $\star$-product which 
controls the dynamics when Wigner distributions are transformed into Husimi 
distributions \cite{takahashi}, 
a smoothed representation popular in applications.  
It is plain that the Gaussian-Laplacian factors 
$T^{-1} (\partial_x,\partial_p)\equiv   \exp(-\hbar(\partial _{x}^2 +
\partial _{p}^2)/4)$ merely dress the standard star-product into Voros' 
product \cite{voros}, 
\begin{equation}
T (f\star g)=T(f) \bowstar T(g)   ~. 
\end{equation}
Consequently, the Lie algebra of brackets of $\phantom{.}\bowstar\phantom{.}$ 
is isomorphic to the Moyal algebra \cite{moyal} 
(the algebra of brackets of $\star$, i.e.\ 
$\{\{f,g\}\}\equiv f\star g- g\star f$), 
in comportance with the general result on the essential uniqueness of the
Moyal algebra as the one-parameter deformation of the Poisson Bracket algebra
\cite{vey}.

Actually, in Fourier space, this product in its original representation 
(\ref{vorosproduct}) appears simpler than the $\star$-product,
\begin{equation} 
f\bowstar g={1\over 2 \pi \hbar }\int d{\bf r}^{\prime}  
d{\bf r}^{\prime\prime}  ~f(x',p')~g(x'',p'')~\delta (x''-x)
  \delta(p'-p)   \exp \left({i\over \hbar} 
(x''-x')(p'-p'') \right) .
\end{equation}
The phase-space integral is then effectively a two-dimensional $\int dx'dp''$,
 not a four-dimensional one, as the kernel has vanishing support everywhere 
but on the lines $x''=x$, $p'=p$. The triangle whose doubled area 
multiplies $-i/\hbar$ in the exponent is now a phase-space {\em right} triangle 
$({\bf r}'',{\bf r}',{\bf r})$, with its side ${\bf r}  - {\bf r'}$ 
horizontal, and its side ${\bf r}  - {\bf r''}$ vertical: 

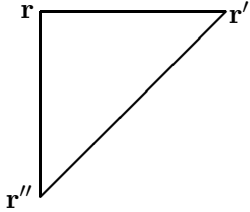
\begin{picture}(100,90)  \thicklines
\put(20,90){\line(1,0){70}}
\put(20,90){\line(0,-1){70}}
\put(20,20){\line(1,1){70}}
\put(15,90){\makebox(0,0)[cc]{{\bf r}}} 
\put(12,20){\makebox(0,0)[cc]{{\bf r}$''$}} 
\put(97,90){\makebox(0,0)[cc]{{\bf r}$'$ }} 
\end{picture}

The triple product is then seen to be actuating shifts on a rectangular lattice,
\begin{eqnarray} 
(f\bowstar g) \bowstar h   &=&
{1\over (2 \pi \hbar)^2 }\!\int \! d{\bf r}^{\prime}  
d{\bf r}^{\prime\prime}  d{\bf r}^{\prime\prime \prime} 
f(x',p') g(x'',p'')h(x''',p''')~\times \\
&\times&\delta (x'''-x) \delta(p'-p) ~\exp 
\left( {i\over \hbar} (    x'(p''-p')+ x''(p'''-p'') +x'''(p'-p''') )\right) .
\nonumber
\end{eqnarray} 
The phase is a cyclic expression with no memory of the order of association,
which thus proves associativity for this product, 
$(f\bowstar g)\bowstar h= f\bowstar (g\bowstar h)$. Pictorially,
the phase is the area of the entire encompassing rectangle with 
diagonal ${\bf r}'''-{\bf s}$, minus the area of the rectangle with diagonal 
${\bf r}'-{\bf r}''$; which is also equal to the {\em  sum} of the areas of 
the rectangles with diagonals ${\bf s}'-{\bf r}'$, and ${\bf r}'''-{\bf r}''$, 
respectively. (In general, it is not twice the area of the triangle 
$({\bf r}', {\bf r}'',{\bf r}''')$.)

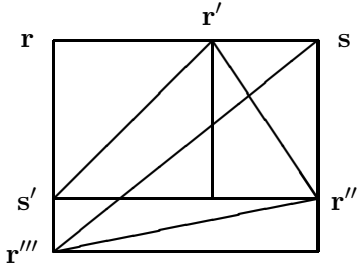
\begin{picture}(130,90)  \thicklines
\put(20,0){\line(1,0){100}}
\put(20,0){\line(0,1){80}}
\put(20,80){\line(1,0){100}}
\put(120,0){\line(0,1){80}}
\put(20,20){\line(1,0){100}} 
\put(80,20){\line(0,1){60}}
\put(20,20){\line(1,1){60}}
\put(20,0){\line(5,1){100}}
\put(20,0){\line(5,4){100}}
\put(80,80){\line(2,-3){40}}
\put(10,0){\makebox(0,0)[cc]{{\bf r}$'''$ }} 
\put(10,20){\makebox(0,0)[cc]{{\bf s}$'$}}
\put(10,80){\makebox(0,0)[cc]{{\bf r}}}
\put(80,90){\makebox(0,0)[cc]{{\bf r}$'$}}
\put(130,80){\makebox(0,0)[cc]{{\bf s}}}
\put(130,20){\makebox(0,0)[cc]{{\bf r}$''$}}
\end{picture}

The construction for an n-tuple $\bowstar$-product follows simply, 
\begin{equation}
{1\over (2 \pi \hbar)^n }\int d{\bf r}_1 ... d{\bf r}_n ~f_1(r_1) ...
f_n(r_n)~\delta (x_n-x) \delta(p_1-p) \exp 
\left( {i\over \hbar} \sum^n_{m=1} x_m (p_{m+1}-p_m) \right) ,
\end{equation}
where $p_{n+1}$ is defined as $p_1$. One may note the effective 
nearest-neighbor interaction in the chain suggested.

{\bf \S 4.} ~A superspace generalization of the star-product was introduced 
in ref \cite{fz}, 
(to codify the graded extension of Moyal's algebra introduced in 
ref \cite{ffz}), 
\begin{equation}
(1+ \hbar \stackrel{\leftarrow }{\partial }_{\theta}
\stackrel{\rightarrow }{\partial }_{\theta} )~\star ~ \equiv 
~\diamond ~\star~.
\end{equation}
Here, $\theta$ is the superspace Grassmann variable (nilpotent, and commuting 
with the phase-space variables): the extended star-product is then a 
direct product of the conventional piece with a superspace factor 
$1+ \hbar \stackrel{\leftarrow }{\partial }_{\theta}
\stackrel{\rightarrow }{\partial }_{\theta}$. Thus, the above extended product 
could have been alternatively written as 
\begin{equation}
e^{\hbar \stackrel{\leftarrow }{\partial }_{\theta}
\stackrel{\rightarrow }{\partial }_{\theta} }~\star ~.
\end{equation}
Hence, it can also be rewritten \cite{fradkin} 
as the evocative form, 
\begin{equation}
e^{{\frac{i\hbar }{2}}(\stackrel{\leftarrow }{\partial }_{x}
\stackrel{\rightarrow }{\partial }_{p}-\stackrel{\leftarrow }{\partial }_{p}
\stackrel{\rightarrow }{\partial }_{x}) +
\hbar \stackrel{\leftarrow }{\partial }_{\theta}
\stackrel{\rightarrow }{\partial }_{\theta}}~.
\end{equation}

Nevertheless, the original form displays associativity more readily, since 
the factor acting on the Grassmann structure is patently associative,
\begin{equation}
\left( A     \diamond B \right) \diamond 
C= A \diamond 
\left( B \diamond 
C \right),
\end{equation}   
acting on 
1d bosonic superfields $A(\theta) =a +\theta \alpha $, 
$~B(\theta) =b +\theta \beta $, so that 
\begin{equation}
A\diamond B=ab +\hbar \alpha \beta +\theta(\alpha b+ a\beta).
\end{equation}
Note the loose analogy to complex multiplication $\overline{z}_1 z_2$.
Even though this analogy cannot rise to an isomorphism, as evident from 
its noncommutativity and longer products such as the above, still, 
it turns out to be useful for actual 
evaluation of products in collecting the Grassmann even and odd terms in 
the answer. The symmetry of this product is further displayed by 
setting $\hbar=1$ and considering standard Grassmann Fourier transforms 
from bosonic to fermionic superfields,
$\tilde{A}(\theta)=\int d\phi (1+\phi\theta)~ A(\phi)
=\alpha+ \theta a $:
\begin{equation}
A\diamond B= \tilde{ A} \diamond \tilde{B}.
\end{equation}

\noindent{\Large{\bf Acknowledgments}} 

Appreciation of discussions with T Curtright and D Fairlie is recorded.
This work was supported in part by the US Department of Energy, 
Division of High Energy Physics,   Contract W-31-109-ENG-38. 


\end{document}